# Application of Computer Technology in Financial Investment

Xinye Sha[a, *]

[a]Graduate School of Arts and Sciences, Columbia University, NY 10027, America
xs2399@columbia.edu

**Abstract:** In order to understand the application of computer technology in financial investment, the author proposes a research on the application of computer technology in financial investment. The author used user transaction data from a certain online payment platform as a sample, with a total of 284908 sample records, including 593 positive samples (fraud samples) and 285214 negative samples (normal samples), to conduct an empirical study on user fraud detection based on data mining. In this process, facing the problem of imbalanced positive and negative samples, the author proposes to use the Under Sampling method to construct sub samples, and then perform feature scaling, outlier detection, feature screening and other processing on the sub samples. Then, four classification models, logistic regression, K-nearest neighbor algorithm, decision tree, and support vector machine, are trained on the processed sub samples. The prediction results of the four models are evaluated, and the results show that the recall rate, Fl score, and AUC value of the logistic regression model are the highest, indicating that the detection method based on computer data mining is practical and feasible.

## 1 Introduction

Since the 1950s, the development of financial information has led to major changes in banking, securities, investment and other related areas, and has promoted the development quickly on various financial products. This change is often called "two revolutions of Wall Street". Financial IT professionals have become the most sought-after skills on Wall Street. After joining the WTO, financial markets will gradually become global, and the integration of skills that understand finance and information technology is important. The importance and necessity of using computer technology in financial investment is a course with deep knowledge and a high foundation of practical application. Computer technology is not only an important tool for learning mathematical theory of finance, but also an important goal of financial analysis and decision making[1-2].

The financial industry generates a lot of data in its daily business, and using current computer technology and database systems can perform tasks such as accessing data, querying , and statistics. Meanwhile, using computer technology and data mining techniques can identify hidden patterns in this large amount of data and reduce the risk of financial institutions.

## 2 Literature Review

Due to the rapid development of computer network technology, the concept of "Internet+" has become popular in recent years and has changed many industries, especially the combination of the Internet and financial capital. Internet money is increasing but users are being cheated because it happens all the time. Therefore, how to detect fraudulent users in advance, successfully attack them, and prevent risks has become a problem that needs to be solved urgently.

To solve this problem, foreign researchers have studied many models and methods based on machine learning to detect fraud from a large number of data changes. Many researchers have tried to use different models to study this problem. Lu ZHANG uses a neural network model to identify credit cards from multiple transaction records, and business detection methods based on this model are used for credit card fraud at Mellon Bank in the United States [3]. Pacher, C.attempts to use Bayesian belief networks for business fraud. After comparing the recognition of Bayesian belief network with previous neural network models, it was found that Bayesian belief network is more reliable and faster than neural network. However, this model has the disadvantage that it is more complex than neural networks and is easy to overfit [4]. Therefore, given the current scams in consumer technology financial investments, the author recommends protection and control from our advice, that is, to improve the level of financial control on the Internet, use financial tools to prevent fraud, and create order on the Internet. financial system. business.

## 3 Methods

### 3.1 Experimental Method Design

This data comes from user logs of some online payment systems and was collected and shared during a research collaboration between Worldline and the machine learning team at ULB (University Libre de Bruxelles) . This is the task of monitoring the distribution of machine learning because every transaction has a label (ie whether it is fraudulent or not). This test will be the first to do preliminary data [5], including the use of pre-engineered low-level models, and then engineering features such as measurement parameters, measurement parameters, and measurements the measure. Four supervised machine learning algorithms were used to train and evaluate the model, and several benchmarks were used to evaluate the performance of each model. A typical test method for fraud detection is shown in **Figure 1**.

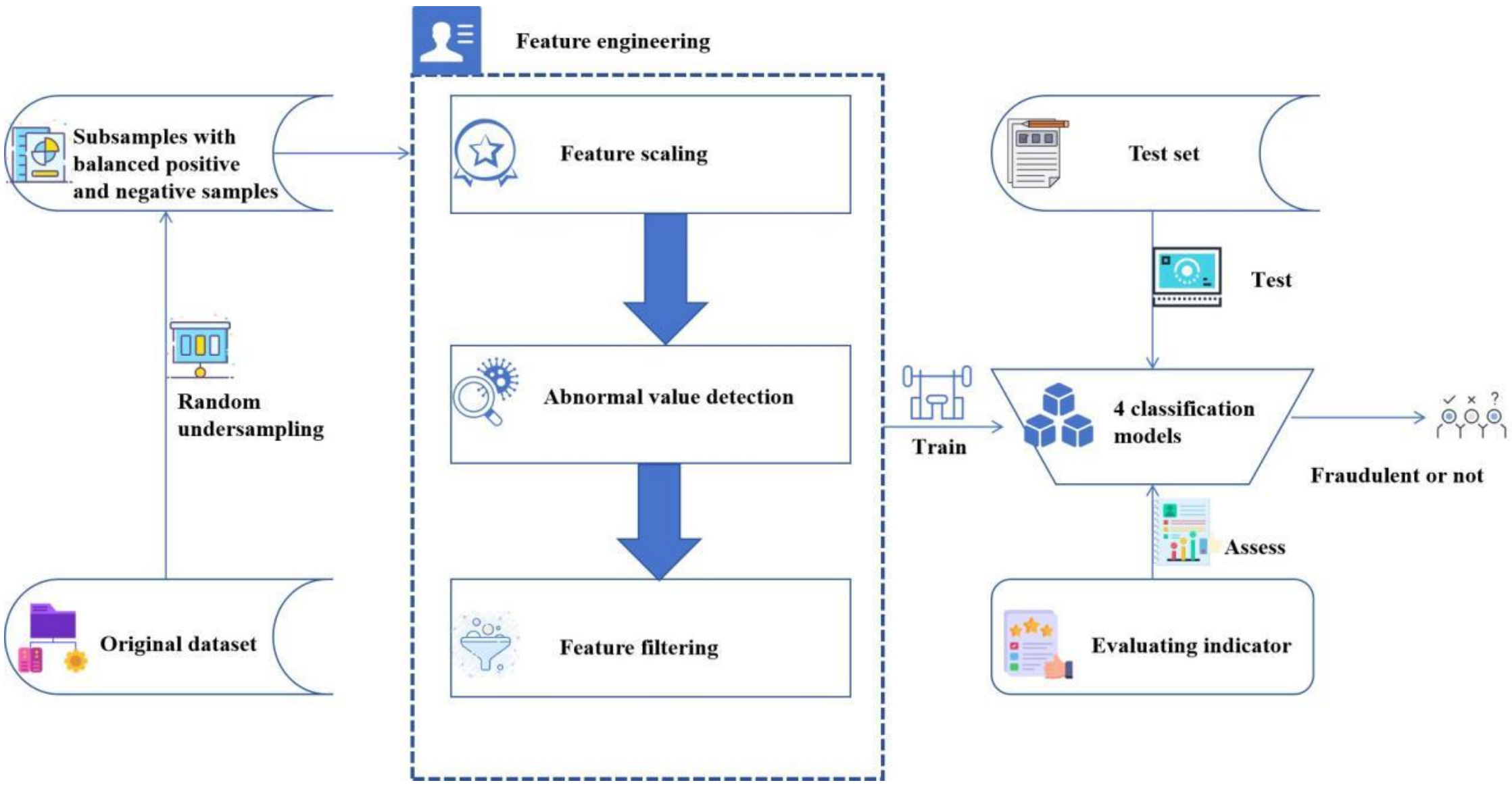

**Figure 1 Experimental flowchart**

## 3.2 Data preprocessing

### 3.2.1 Data preparation

The first thing we need to do is to have a basic understanding and knowledge of datasets. This dataset contains online payment transaction records from Worldline Company in October 2020, with a total of 284908 sample records, including 593 positive samples (fraud samples) and 285214 negative samples (normal samples). As shown in **Figure 2**, it can be observed that the dataset is highly imbalanced, with fraud samples accounting for only 0.219% of all transactions[6].

In addition, the dataset contains a total of 30 variables, all of which are numerical variables. Among them, 5 variables were anonymized due to confidentiality issues, namely V1, V2, V3, V4, V5. In addition, there are 2 variables, namely transaction time (the number of seconds passed between this transaction and the first transaction in the dataset) and transaction amount. The last variable is a category label, which is a binary value that can have a value of 0 (non fraudulent) or 1 (fraudulent). There are no null values in the values of each variable, so there is no need for null value processing.

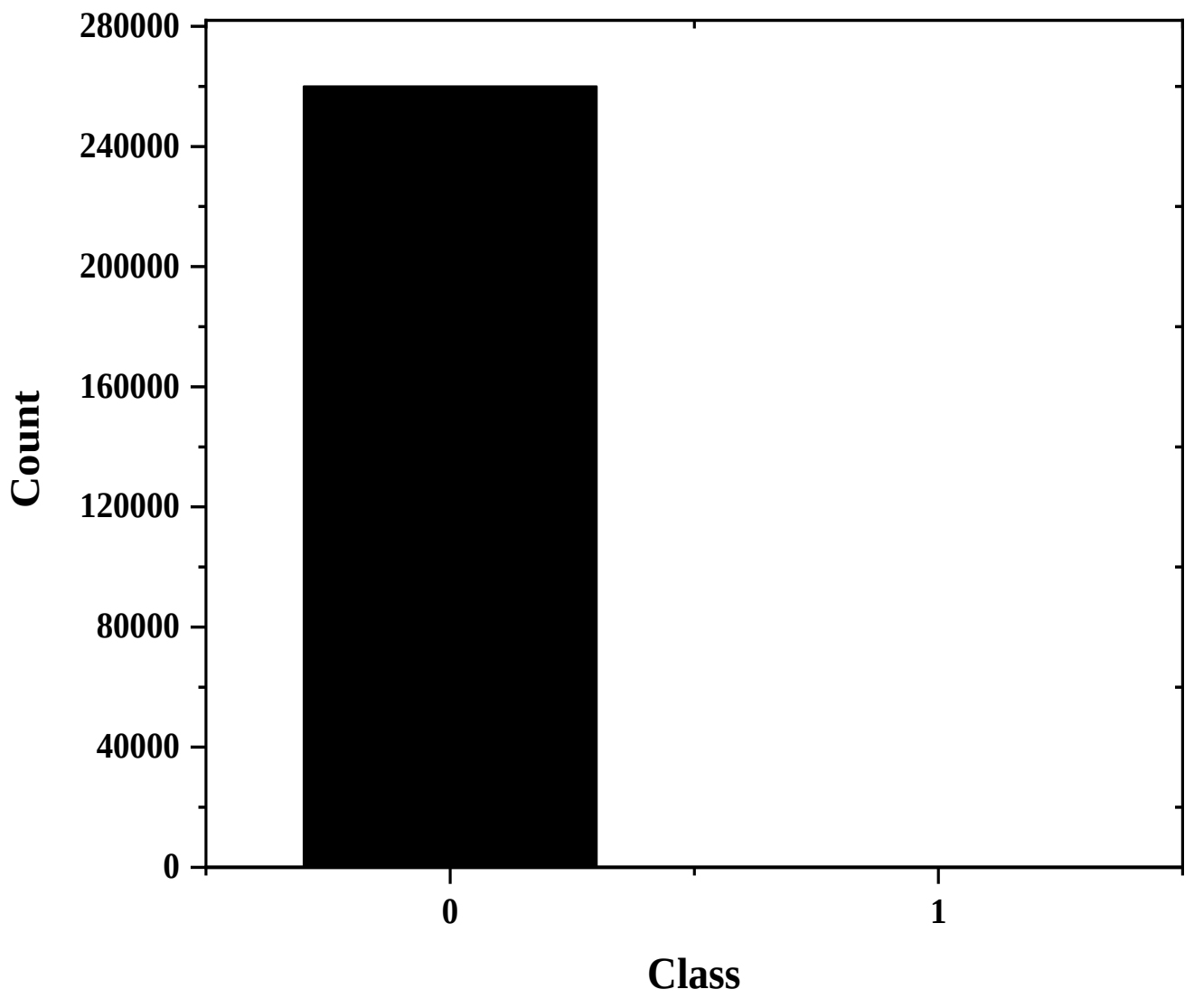

**Figure 2 Distribution of positive and negative sample sizes**

### 3.2.2 Random undersampling

Considering the simplicity of operation and the avoidance of model overfitting, the author chooses to use a random undersampling method to construct a 3/3 balanced subsample. In order to verify the effectiveness of random undersampling, we can compare the feature correlation matrices before and after sample processing. **Figure 3** shows the correlation matrix of the original dataset, and it can be seen that almost all variables in the original dataset are uncorrelated. **Figure 4** shows the feature correlation matrix of the sub samples constructed through undersampling. It can be seen that there is a strong negative correlation between V3, V4, and V5 in the sub samples and the classification results; V2, V4, and V5 are positively correlated with the classification results. This proves the effectiveness of random undersampling through the correlation between features [7].

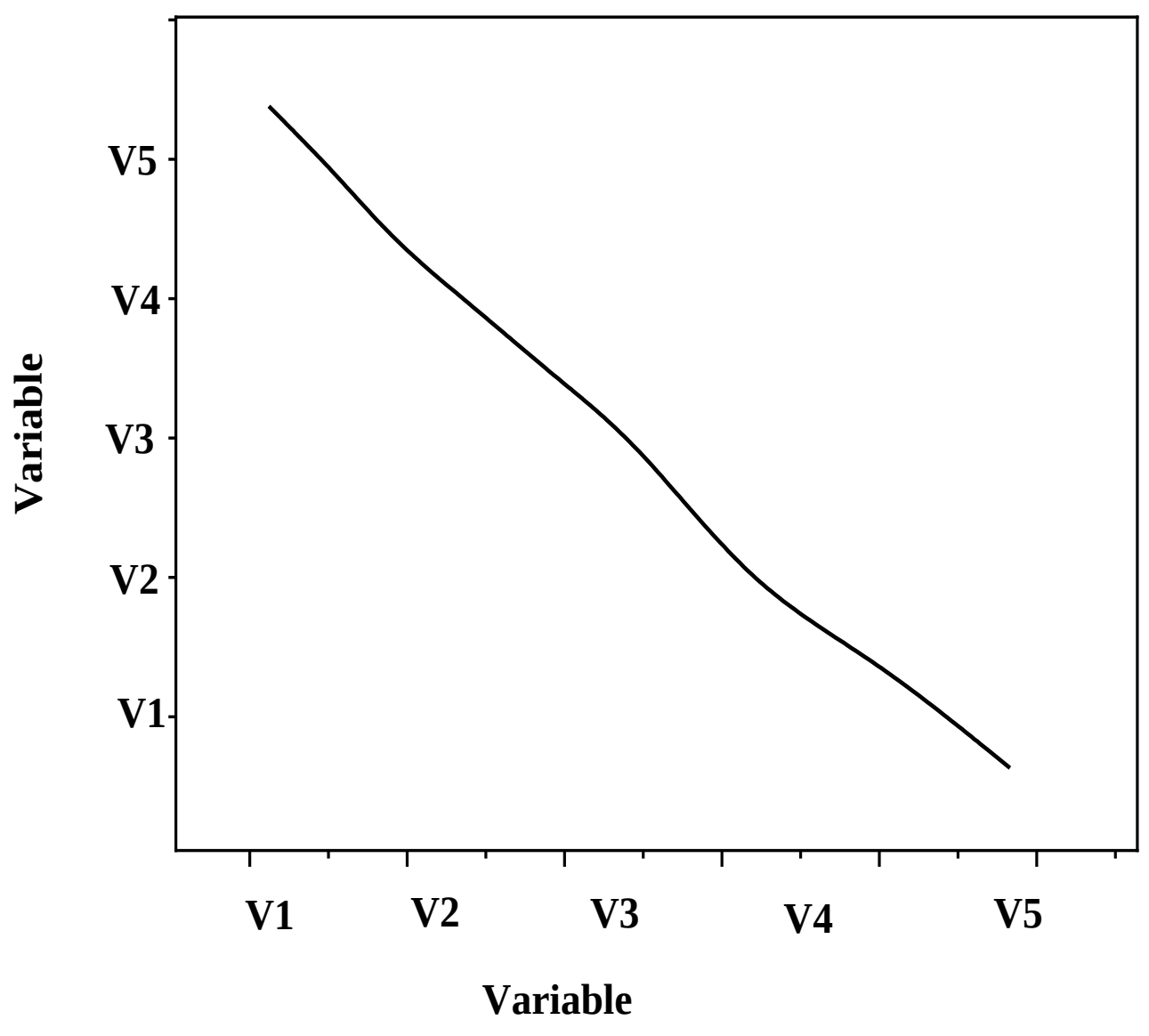

**Figure 3 Correlation Matrix of the Original Dataset**

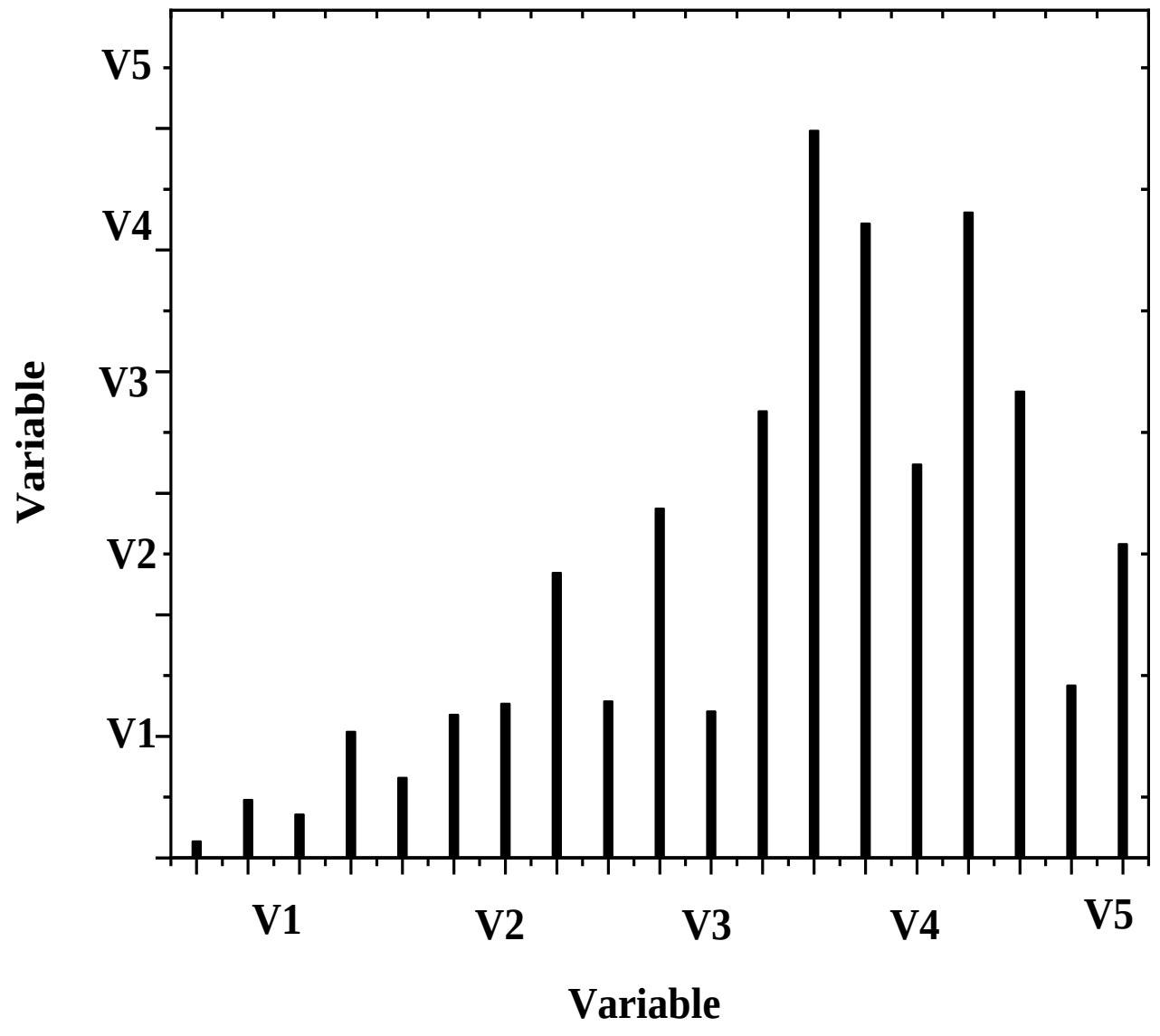

**Figure 4 Correlation Matrix of Subsamples**

### 3.2.3 Feature scaling

Feature Scaling refers to the method of quantifying the values of different features into the same interval. Simply put, it involves placing features that originally had significant differences in numerical ranges due to different units into the same numerical interval [9]. There are three commonly used methods for feature scaling:

(1) The minimum maximum normalization is shown in equation (1), where m 'represents the scaled value, m represents the original value, and min (m) represents the small interval of the feature dimension. When min (m) or max (m) is an outlier or outlier, the distribution of the scaled data will be extremely uneven, which will affect subsequent data processing.

$$m' = \frac{m - \min(m)}{\max(m) - \min(m)} \,)(1)$$

(2) Mean normalization: See equation (2), where m 'represents the scaled value, m represents the original value, min (m) represents the minimum value of the feature dimension, max (m) represents the maximum value, and average (m) represents the average value. This method can scale all values to between [-2, 2] and have a mean of 0.

$$m' = \frac{m - \mathrm{av}erage(m)}{\max(m) - \min(m)} (2)$$

(3) Standardization: See equation (3), where is the mean value of the feature dimension and α is the standard deviation. It can be ensured that all dimensions of the input data follow a normal distribution with mean 0, variance 2, and so on.

$$m' = \frac{m - m}{\alpha} (3)$$

**3.2.4 Abnormal value detection**

Outliers refer to data values that differ significantly from the overall structure, characteristics, attributes, etc. We need to examine the correlation matrix of each feature in order to understand which features have a high positive or negative correlation with the classification results [8]. From **Figure 4**, we can see that the features with the highest correlation with the classification results are V4, V3, and V5. Therefore, we will identify extreme outliers from these three features and remove them.

**3.3 Training and Testing of Models**

In this section, we will train and test four types of classifiers and determine which one will be more effective in detecting user fraud applications. The four classification algorithms used are logistic regression, K-nearest neighbor algorithm [10], support vector machine, and decision tree, which are commonly used and classic mature algorithms in binary classification problems.

**4 Result analysis**

### 4.1 Evaluation indicators

The various indicators are commonly used to evaluate binary classification models in machine learning, including precision, return value, the F1 value, the horizontal and vertical lines of the ROC curve and the AUC value of the ROC curve are false positive rate (FPR) and true positive rate (TPR).

### 4.2 Analysis of experimental results

Firstly, calculate the accuracy, recall, and F1 score of each model's prediction results on the test set. After sorting, the results are shown in **Table 1**. In this experimental context, we will pay more attention to the relevant predictive indicators of fraudulent samples (with category label 1). In this case, the accuracy represents how many of the predicted fraudulent samples are truly fraudulent, and the recall represents how many of all fraudulent samples can be predicted by the model. Comparison shows that there is not much difference in the accuracy of the four models in predicting fraudulent samples, but the logistic regression model has the highest recall and F1 score, followed by support vector machine. According to the previous evaluation indicators, we know that the higher the recall rate and F1 value, the better the model performance. That is to say, under these indicators, the logistic regression model predicts the best performance.

**Table 1: Values of various indicators for the four models**

| model | Accuracy | recall | F1 value |
|---|---|---|---|
| logistic regression | 0.86 | 0.83 | 0.86 |
| K-nearest neighbor algorithm | 0.85 | 0.76 | 0.80 |
| Support vector machine | 0.87 | 0.81 | 0.84 |
| decision tree | 0.84 | 0.73 | 0.77 |

Next, we evaluate and compare the ROC curve and take the AUC values of the four models. As shown in **Figure 5**, the ROC curve of the four models can reflect the difference in sensitivity (FPR) and precision (TPR) of the model. Therefore, the closer the curve is to the upper left, the smaller the horizontal FPR and the larger the vertical TPR, the better the model. You can see that the logistic regression and support vector machine curves are closest to the upper left.

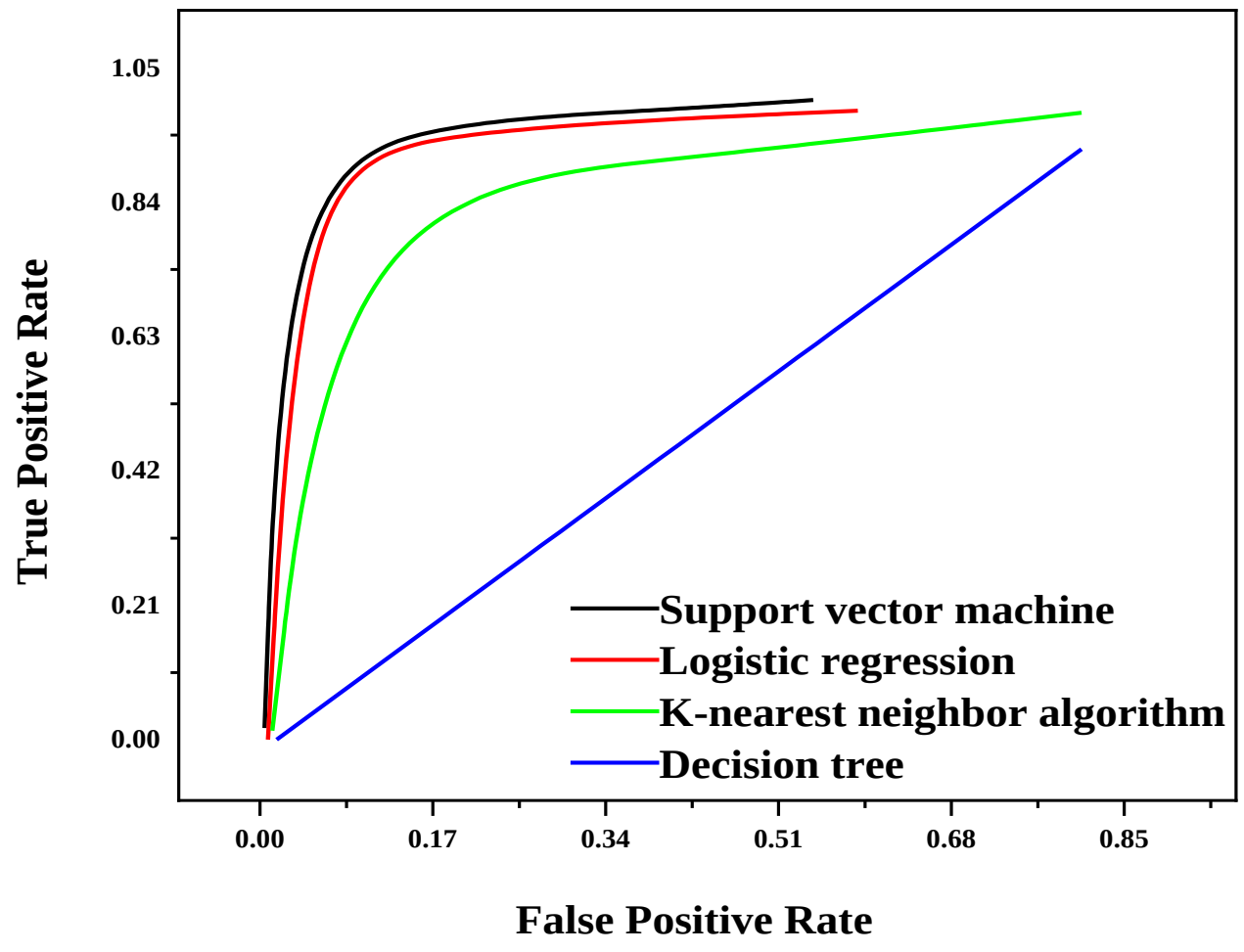

**Figure 5 ROC curves of four models**

Comparing the AUC values of the four models as shown in **Table 2**, the logistic regression model still has the highest AUC value, followed by the support vector machine. The higher the UC value, the better the model is able to separate between positive and negative samples. Therefore, the logistic regression model and the support vector machine model still perform well in this test.

**Table 2 AUC values for four models**

| model | AUC value |
|---|---|
| logistic regression | 0.989 |
| K-nearest neighbor algorithm | 0.925 |
| Support Vector Machine | 0.975 |
| Decision Tree | 0.918 |

## 5 Conclusion

This paper combines literature research and empirical research based on data mining to explore this issue, and obtains the following research results: Empirical research on user fraud detection based on data mining on real online payment transaction datasets. In this process, facing the problem of imbalanced positive and negative samples, it is proposed to use the Under Sampling method to construct sub samples. Then, four classification models, namely logistic regression, K-nearest neighbor algorithm, decision tree, and support vector machine, were trained on the sub samples processed by feature engineering. The prediction results of the four models were evaluated, and the results showed that the logistic regression model had the highest recall rate, F1 score, and AUC value, indicating that the logistic regression model performed the best in predicting in this experimental dataset. This also indirectly proves that the user fraud detection method based on computer technology data mining is practical and feasible.